\documentstyle{mn}
\input psfig

\newif\ifAMStwofonts
\AMStwofontstrue

\def\lesssim{\mathrel{\hbox{\rlap{\hbox{\lower4pt\hbox{$\sim$}}}\hbox{$<$}}}}
\def\gtrsim{\mathrel{\hbox{\rlap{\hbox{\lower4pt\hbox{$\sim$}}}\hbox{$>$}}}}

\def\aj{{AJ}}			
		
\def\apj{{ApJ}}			
		
\def\apjs{{ApJS}}

\def\aap{{A\&A}}

\def\mnras{{MNRAS}}

\def\nat{{Nature}}

\title[Faint galaxies close to QSOs with damped Lyman-$\alpha$ absorption systems]{Faint galaxies close to QSOs with damped Lyman-$\alpha$ absorption systems}

\author[Alfonso Arag\'on-Salamanca et al.]{Alfonso Arag\'on-Salamanca,$^1$ Richard S. Ellis,$^1$ and Kieran S. O'Brien$^2$\\
$^1$Institute of Astronomy, Madingley Road, Cambridge CB3 0HA, England\\
$^2$Royal Greenwich Observatory, Madingley Road, Cambridge CB3 0EZ, England
}

\date{Accepted ---.
      Received ---;
      in original form ---}

\begin{document}

\label{firstpage}

\maketitle

\begin{abstract}

We have obtained very deep near-infrared images in the fields of 10
QSOs whose spectra contain damped Lyman-$\alpha$ absorption (DLA)
systems with $1.7<z_{\rm abs} <2.5$.  The main aim of our investigation
is to provide new constraints on the properties of distant galaxies
responsible for the DLA absorption. After subtracting the point spread
function associated with the QSO light, we have detected galaxies very
close to the QSO line of sight (projected distance
$1.2$--$1.3\,$arcsec) in two fields.  These sources therefore represent
promising candidate galaxies responsible for the DLA absorption. Placed
at the absorber's redshift, the impact parameter is $10h_{50}^{-1}\,$kpc and
the luminosity is close to $L_K^{\ast}$. Such parameters are consistent
with the hypothesis, verified for metallic systems at lower redshift,
that slowly-evolving massive galaxies produce at least some of  the
absorption systems of high column density in QSO spectra out to beyond
$z\simeq2$.  In addition to detecting these candidate DLA galaxies, the
radio-loud QSOs in our sample show a significant excess of  sources on
larger scales ($\theta \simeq 7\,$arcsec); this excess is not  present
in the radio-quiet QSO sightlines. We calculate that such an excess
could be produced by luminous galaxies in the cores of clusters
associated with radio-loud QSOs. Both results confirm that deep imaging
of selected QSOs can be a powerful method of finding samples of likely
$z\simeq2$ galaxies.  Follow-up near-infrared spectroscopy is required
to secure galaxy redshifts and star formation rates, while deep HST
imaging can determine sizes and morphologies, providing valuable
information on galaxy properties at large look-back times.

\end{abstract}


\section{Introduction}

Considerable progress has been made in recent years in understanding
the evolutionary behaviour of starlight from distant field galaxies.
Systematic redshift surveys (Ellis et al. 1995, Lilly et al. 1995) have
provided large datasets from which the evolution of the luminosity
function of galaxies has been directly determined to redshifts
$z\simeq1$. Such studies are complemented by smaller, but independent
surveys based on the absorbing properties galaxies present to
background luminous QSOs (Bahcall \& Peebles 1969).  Following the
pioneering work of Bergeron and Boiss\`e (1990), Steidel \& Dickinson
(1995) have identified those galaxies responsible for the MgII
absorption seen in the sightlines to 68 QSOs and have thus derived
luminosity functions and rest-frame colours for the absorbing field
galaxies at a mean redshift of $z\simeq0.7$.

It is gratifying that both techniques, viz surveys that select galaxies
by their {\it emission\/} and those via their {\it absorption}, give
consistent results for the evolutionary behaviour of massive galaxies
to $z\simeq1$ (Steidel \& Dickinson 1995, Ellis 1995). Surprisingly
little evolution is seen in both the colours and luminosities of such
galaxies indicating they were already in place at large look-back
times. The principal difference between the techniques lies in
understanding the origin and fate of the numerous star-forming dwarfs
seen in the redshift surveys (Ellis 1995); such galaxies do not appear
to be prominent in the metallic absorber-selected samples.

The slow evolution of massive galaxies to $z\simeq1$ has important
implications for the likely detection of similar systems at higher
redshift, particularly for early-type disk galaxies whose star
formation rates are expected to increase slightly with redshift.
Limited field spectroscopy has been performed beyond $z=1$ (Cowie, Hu
\& Songaila 1995) but QSO absorption samples are available in
abundance. Indeed, some constraints on the nature of galaxies
responsible for high $z$ metallic systems are available from
statistical studies of a few MgII systems (Steidel et al. 1995, private
communication) and a sample of more distant CIV absorbers
(Arag\'on-Salamanca et al. 1994).  Although no spectroscopic
identifications have yet been made for these systems, the implied
luminosities and impact parameters are consistent with continuation of
the trends found at lower redshift.

Of particular interest in this regard is the nature of the material
which produces the damped Lyman alpha absorption (DLA, Wolfe et al.
1986).  Conventional wisdom holds that such systems arise infrequently
when the disk of an absorbing galaxy presents a small impact parameter
to the QSO.  Statistical analyses indicate the abundance of MgII, CIV
and DLA systems is consistent with a single population of massive
galaxies whose volume density approximates that seen today, the
principal difference arising in the cross-section. If this is the case,
deep imaging in good seeing will be required to isolate the galaxies
responsible for the DLA systems. Although this explanation is not
accepted universally (see, e.g., Yanny \& York 1992), any constraints
on the properties of galaxies close to the QSO sightline where a DLA
absorber may be present will be of interest.

This paper begins to explore the possibilities observationally. In
Section 2 we present new infrared imaging observations of 10 QSO
sightlines each chosen to contain a damped Lyman alpha absorption.
Section 3 analyses the distribution of faint sources found as a
function of projected distance from the QSO, and discusses two
promising cases where the DLA absorber may have been found. The
implications are briefly discussed and Section 4 presents our
conclusions.

\section{The data} 

The QSOs for this study have been selected from a compilation of
confirmed damped Lyman-$\alpha$ systems kindly provided by Max Pettini
(Table~1). We selected all the DLA systems with $1.5 < z_{\rm abs} <
2.5$ and $z_{\rm abs} \ll z_{\rm em}$ in the $8^{\rm h} < {\rm R.A.} <
15^{\rm h}$ right ascension range. We imaged 10 QSOs out of the 11 in
the Pettini et al. list that fulfill these criteria, the exception
being Q1409+095 which was not observed due to the limitations in
available telescope time.  Out of the 10 observed QSOs, 6 are optically
selected with no reported radio emission; the remaining 4 are radio
loud  (Hewitt \& Burbidge 1993). We shall refer to these as the `radio
quiet' and `radio loud' samples respectively.

\begin{table*}
\centering
\begin{minipage}{14.5cm}
\begin{center}
\caption{The QSO sample.}
\begin{tabular}{l c c c c c c c}
\hline
{QSO} & 
{$z_{\rm em}$} & 
{$z_{\rm abs}$} &
{$z_{\rm abs}$} &
{Exp. time} & 
{Seeing} & 
{$K_{\rm lim}$} & 
{$\theta_{\rm min}$} \\
{} & 
{} & 
{DLA} &
{Metals} &
{(seconds)} & 
{FWHM ($^{\prime\prime}$)} & 
{(mag)} & 
{($^{\prime\prime}$)}\\
\hline
0836$+$113 & $2.696$ & $2.465$ & $0.787^{\rm a}$ & $12960$ & $0.9$ & $21.7$ & 0.6 \\
0841$+$129$\,^{\rm r}$ & $>2.51^{\rm b}$ & $1.861, 2.375, 2.477$ &   & $12480$ & $1.0$ & $21.7$ & 0.9 \\
1100$-$264$\,^{\rm r}$ & $2.145$ & $1.839$ & $0.359, 1.219$ & $6480$ & $1.0$ & $21.3$ & $2.0$ \\
1136$+$122 & $2.894$ & $1.789$ & $0.317, 2.074$ & $6480$ & $0.9$ & $21.3$ & $1.0$ \\
1151$+$068 & $2.762$ & $1.774$ & $0.684, 1.819, 2.024$ & $6480$ & $1.1$ & $21.3$ & $1.6$ \\
1215$+$333$\,^{\rm r}$ & $2.606$ & $1.999$ & & $11820$ & $1.2$ & $21.7$ & $1.2$ \\
1223$+$178 & $2.936$ & $2.466$ & & $12960$ & $0.9$ & $21.7$ & $1.4$ \\
1244$+$347 & $2.500$ & $1.860$ & & $6780$  & $0.9$ & $21.3$ & $1.5$ \\
1331$+$170$\,^{\rm r}$ & $2.081$ & $1.773$ & $0.745, 1.328, 1.446$ & $8400$ & $0.9$ & $21.4$ & $2.0$ \\
1406$+$123 & $2.970$ & $2.248$ & & $11280$ & $1.0$ & $21.7$ & $1.4$ \\
\hline
\end{tabular}
\end{center}

$^{\rm r}${Radio-loud. }

$^{\rm a}${Only confirmed metallic absorption systems at $z^{\rm metals}_{\rm abs} \ne z^{\rm DLA}_{\rm abs}$ are listed. For multiple systems, the mean redshift is given (from Junkkarinen, Hewitt \& Burbidge 1991; and
Hewitt \& Burbidge 1993). }

$^{\rm b}${0841$+$129 is a BL Lac object. Its emission redshift
is a lower limit determined from the highest redshift absorption system detected in its spectrum.  }

\end{minipage}

\end{table*}

The QSOs were imaged in the $K$-band using the $256\times256$ InSb
infrared camera IRCAM3 on the 3.8$\,$m UK Infrared Telescope on Mauna
Kea during the nights of February 24th--27th 1995. All nights were
clear and photometric, with good seeing conditions
($0.5^{\prime\prime}$--$1.2^{\prime\prime}$).  The pixel scale was
$0.3^{\prime\prime}\,$pixel$^{-1}$.  Details of the infrared exposures
are given in Table~1. Exposure times were chosen to ensure the
detection of galaxies as faint as $0.3L_K^\ast$ at $z_{\rm abs}$
assuming no evolution. We adopt $H_0=50\,$km$\,$s$^{-1}$Mpc$^{-1}$ and
$q_0=0.5$ throughout, and $M_K^{\ast} = -25.1$ form Mobasher, Sharples
and Ellis 1993. The $K$-corrections used in this paper followed
procedures described more fully in Arag\'on-Salamanca et al. (1994).

The observing procedure involved numerous dis-registered short
exposures in a $3\times3$ pattern of step size $\simeq
7^{\prime\prime}$. The exact telescope offsets included a random
component of $\simeq 1$--$2^{\prime\prime}$ amplitude in the right
ascension and declination directions which avoids producing images at
exactly the same position with consequent  problems in flat-fielding
(see ahead).  The individual image exposures were 60$\,$s,  composed of
six 10$\,$s background-limited sub-exposures.  Dark frames were
obtained  at frequent  intervals with the same exposure times and
subtracted from the science images.

The median\footnote{Strictly speaking, instead of the median, we used
the Biweight central location estimator, which behaves like the median
when the number of data points is large, but produces a better S/N for
relatively small samples. See Beers et al. (1990) for a detailed
description of this statistical estimator.} of individual images taken
over a~$\simeq 10\,$minute period around a given observation produces a
very good flat-field that yields a uniformity  of better than 3--5
parts in $10^5$ on the final images.

To improve the sampling of the point spread function (PSF), we utilised
the fact that the fields were imaged in many dis-registered positions
with fractional pixel shifts thus using information on smaller spatial
scales. We artificially divided each pixel in four to produce a scale
of $0.15^{\prime\prime}\,$pixel$^{-1}$. The resampled images  were
registered (using sub-pixel shifts) and median combined.  The final
images have well-sampled PSFs with seeings listed in Table~1, and cover
$1^{\prime}\times1^{\prime}$ at their maximum depth.  The images were
reduced using our own software based on the FIGARO package written by
Keith Shortridge.

Photometric calibrations were secured by repeated observations of
standard stars in the UKIRT faint $JHK$ standards list (Casali \&
Hawarden 1992), yielding absolute photometry whose internal accuracy
is better than 0.01$\,$mag {\it r.m.s.}

In order to detect objects as close as possible to the QSO sightlines,
we subtracted a suitably scaled PSF from the QSO images using the
DAOPHOT package in IRAF. To determine the PSF we used bright stars in
the same image when available. Given the relatively small field of
view, this was not always possible. In those cases where no bright star
was present in the field, we determined the PSF from a bright star in
an image with better seeing and convolved it with a two dimensional
Gaussian of the appropriate $\sigma$, ellipticity and orientation  so
that the resulting PSF would have the right shape parameters (as
determined from the QSO image itself). The success of this procedure
depends critically on the QSO brightness, but we achieved residuals
fainter than the limiting magnitude of the images (see ahead) at radii
larger than $0.6$--$2^{\prime\prime}$ from the QSO centre.  Table~1
lists the minimum radius $\theta_{\rm min}$ for which reliable source
detection was possible. The region $\theta<\theta_{\rm min}$ was
patched out and excluded from further analysis.  Our procedure is
arranged so as to  subtract the maximum amount of QSO light by
matching  the PSF parameters to those of the QSO image. This is helpful
in revealing discrete sources ``under'' the PSF rather than residual
QSO ``fuzz''.

Figure~1 shows the images before and after  PSF subtraction.  In the
case of Q1100$-$264, ---the brightest QSO in the sample--- the PSF
subtracted image shows some diffuse residual light which results from
limiting the diameter of the subtracted PSF to $35$ pixels
($5.25\,$arcsec). Note that the bright star in the Q0836+113 shows the
same diffuse structure.

\begin{figure}
\psfig{figure=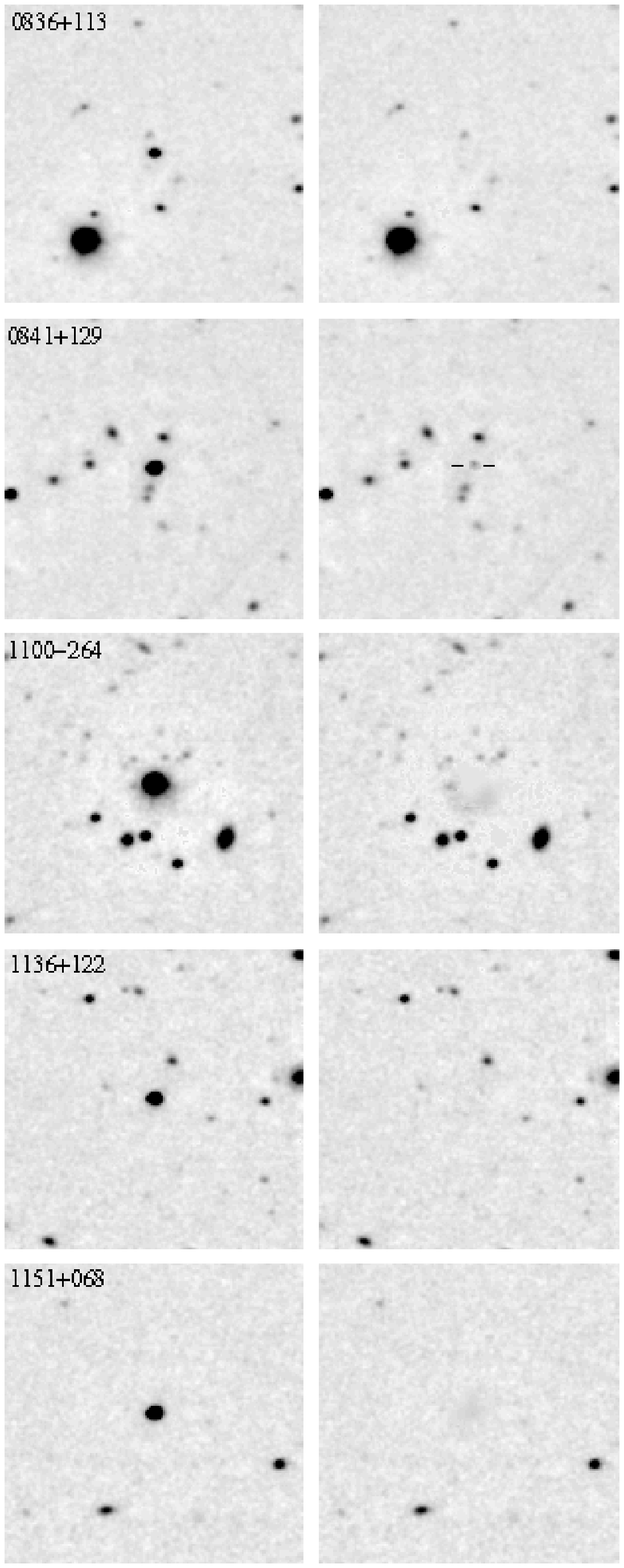,width=170mm}
\begin{minipage}{18.0cm}
\caption{Deep infrared $K$-band images of the QSOs listed in Table~1 
before (left) and after (right) PSF subtraction. 
North is up, East is left. Each frame is $1^\prime\times1^\prime$. 
The two DLA galaxy candidates discussed in Section~3.2 have been marked.}
\end{minipage}
\end{figure}

\begin{figure}
\psfig{figure=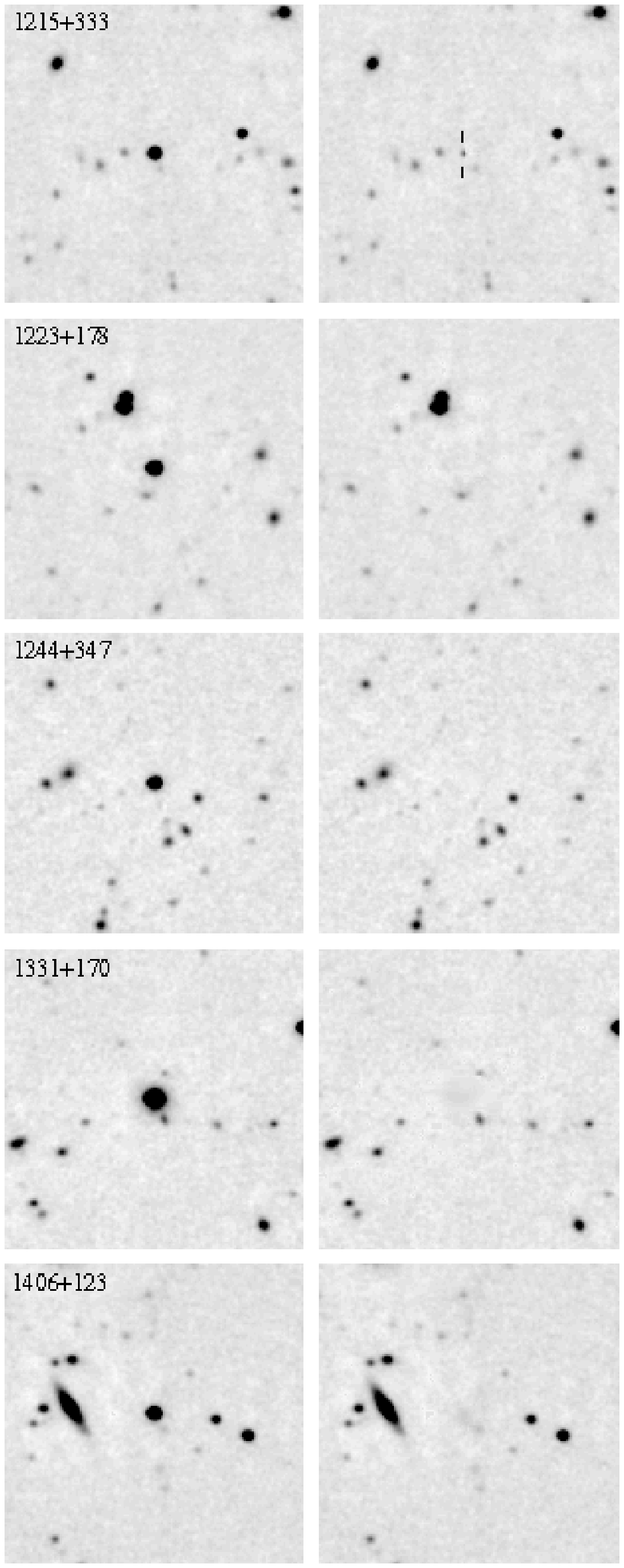,width=170mm}
\end{figure}

In order to construct the $K$-selected galaxy samples, automated object
detection was performed using the APM software (Irwin 1985) in the
STARLINK PISA implementation. An object was detected in the final $K$
images when 6 or more connected pixels had  counts larger than
$3\sigma$ above the background. Adjacent objects were de-blended
following the algorithm described by Irwin (1985).  The software
outputs positions, integrated fluxes, orientation and shape  parameters
for the detected objects.  Integrated photometry in a
$2.5^{\prime\prime}$ diameter aperture was obtained using the PHOTOM
package in STARLINK. This procedure is preferable  to using the APM
magnitudes directly because it provides a local estimation of the sky.
We choose a small aperture for the photometry since, given the good
seeing conditions, this gives an adequate approximation of the total
magnitude while substantially reducing uncertainties from sky
subtraction. 

Photometric random errors were determined  following the procedure
described in Arag\'on-Sa{\-}la{\-}man{\-}ca et al. (1994).  Briefly, we
first divided the total integrations in several sub-exposures, and
examined the scatter in the photometry for individual objects. Second,
we used the sky variance near each object, since for the high $K$-band
background this should be the dominant source of error. Both methods
agreed closely, yielding typical random errors $\simeq0.10$--$0.12$ at
$K\simeq20$ and $\simeq0.20$--$0.25$ at $K\simeq21$. For each field we
adopted that limiting magnitude (Table~1) whose photometric errors are
less than $0.3\,$mag inside the $2.5^{\prime\prime}$ apertures. The
cumulative number counts to these limiting magnitudes show no sign of
incompleteness, and are compatible with the published $K$-band number
counts (Gardner, Cowie \& Wainscoat 1993). The chosen limiting
magnitudes are quite conservative and ensure that no spurious sources
that might have been introduced in the catalogues by the automatic detection
algorithm were kept.

\section{Analysis} 

\subsection{Radial distribution of sources}

Following a similar analysis to that of Arag\'on-Salamanca et al.
(1994), we next studied the distribution of sources as a function of
angular separation $\theta$ from the QSO. Figure~2(a) shows the radial
distribution of sources for the combined sample of all 10 QSOs, both in
numbers to our adopted  $K$ limits and the factor by which the source
density exceeds the Gardner, Cowie and Wainscoat (1993) field number
counts. There is good agreement between the measured number of sources
and that expected from the field number counts at radii larger than
$\simeq10^{\prime\prime}$. At smaller radii there is, however, a
marginal excess.  Interestingly, when the QSO sample is split according
to radio quiet (6 QSOs) and radio loud (3 QSOs and one BL Lac object),
significant differences are seen.  Whereas the radial distribution of
associated  sources for the radio quiet sample is compatible with a
uniform distribution (Figure~2(b)), that for the radio loud QSO fields
shows an excess of sources at radii $\theta\lesssim 7\,$arcsec
(Figure~2(c)). In order to quantify the statistical significance of
these results, we carried out Kolmogorov-Smirnov tests on the
distribution comparing these with a uniform distribution.  For the
combined QSO sample, the deviation from a uniform distribution is
significant only at the $85$\% level. The sources in the radio-quiet
QSO sample show deviations significant at $39$\%, while the sources in
the radio-loud QSO fields are significant at the $99.5$\% level.  
These results are consistent with the error bars in Figure~2. While
the latter significance is still not formally very high given the
sample size, the difference between the two sub-samples suggests that
the excess of sources is unlikely to arise from galaxies associated
with the DLAs whose properties across the entire sample show no such
distinction.

The radial extent of the excess ($\simeq 7\,$arcsec, corresponding to
$\simeq 60\,$kpc at $\langle z_{\rm abs}\rangle=2.06$) implying
cross-sections larger than expected if DLA absorption is produced by
the inner, denser galaxy regions (Steidel et al. 1994). Such radial
distances are also much larger than those at which neutral hydrogen
column densities exceed $10^{21}\,$cm in present day galaxies (Rao \&
Briggs 1993). Of course this does not preclude the possibility that the
excess might be associated with galaxies {\it clustered\/} with the
ones producing the Lyman alpha absorption, but the difference in
behaviour of the radio loud and radio quiet QSO sightlines argues
against this.

\begin{figure}
\psfig{figure=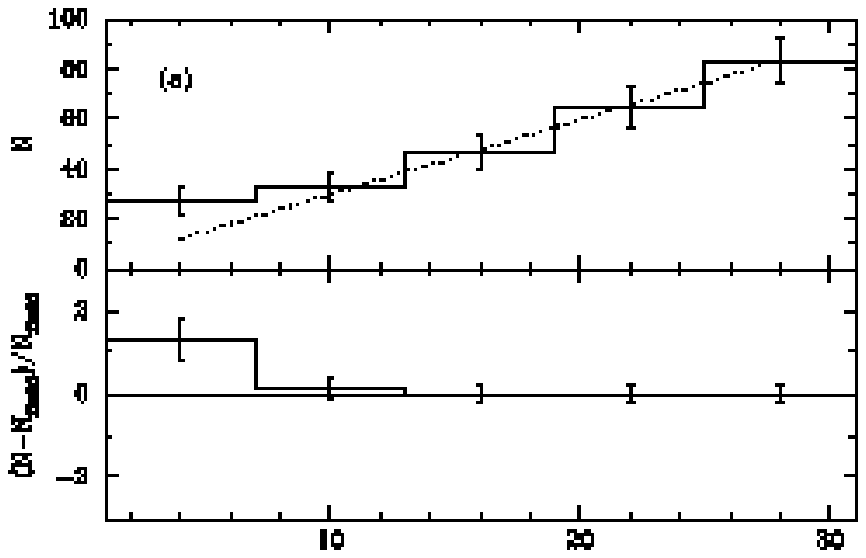,width=80mm,height=59mm}
\vskip 3mm
\psfig{figure=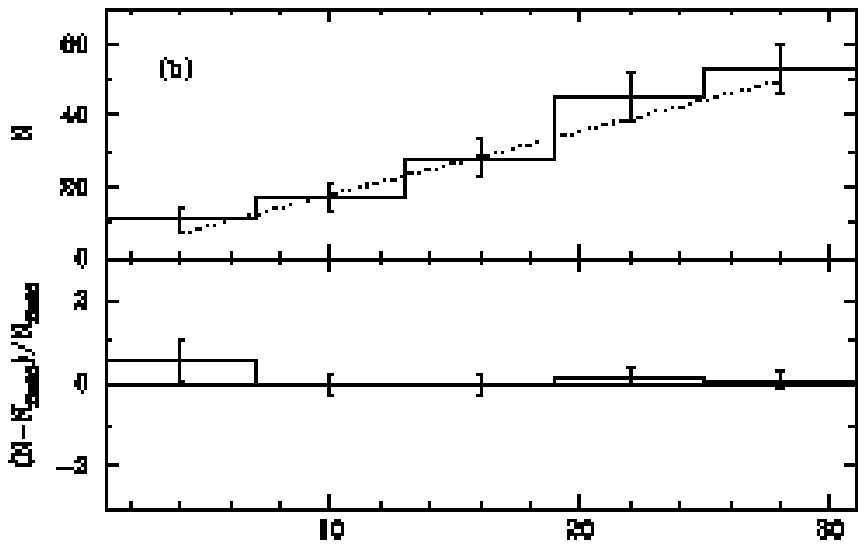,width=80mm,height=59mm}
\vskip 3mm
\psfig{figure=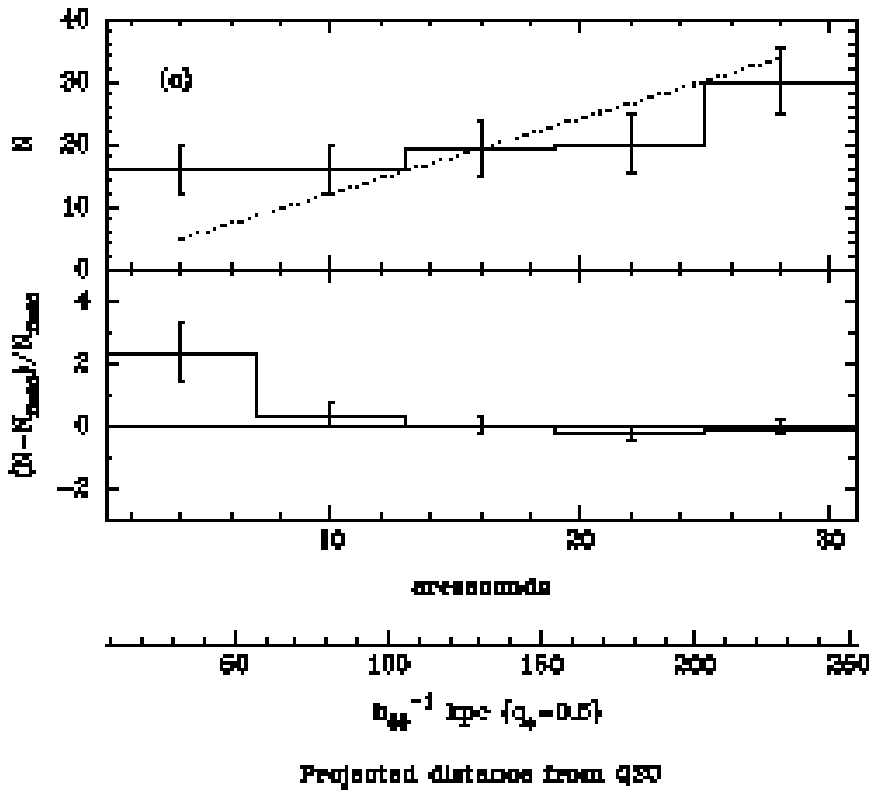,width=81.4mm,height=85mm}
\caption{
{\bf (a)} The summed distribution of faint sources close to the QSOs as a
function of projected radial distance in direct numbers (top) and in
terms of the multiplicative excess over the field counts (bottom). The
dotted line in the upper panel shows the expected field galaxy counts (see text). {\bf (b)} As (a) but for the radio-quiet QSOs.  {\bf (c)} As (a) but
for the radio-loud QSOs.}
\end{figure}

An important consideration at this point is the fact that some of the
QSO spectra show absorption line systems (in particular MgII and CIV)
at redshifts lower than those of the DLA (Table 1). As MgII absorption
lines at $z\lesssim1$ are produced by relatively luminous galaxies with
characteristic distances from the QSO line of sight $\simeq 70\,$kpc
(Bergeron \& Boiss\'e 1991;  Steidel, Dickinson \& Persson 1994,
Steidel \& Dickinson 1995), some of the excess could be associated with
such galaxies.  However, in our sample the number of lower redshift
absorption systems in radio loud QSOs does not differ significantly
from that of radio quiet QSOs, so the presence of an excess in the
former and not in the latter is unexplained.  Moreover, given the lower
redshift of the other metallic absorption systems, we would expect them
to be, on average, at somewhat larger radii form the QSO.  Indeed, for
the one case in our QSO sample where the galaxy producing a low
redshift MgII absorption has been identified (Q1100$-$264, $z_{\rm
abs}^{\rm MgII}=0.359$), the absorbing galaxy is $12\,$arcsec SE from
the QSO (Bergeron \& Boiss\'e 1991), thus outside the $\theta\lesssim
7\,$arcsec region where the excess has been detected.  Taken
collectively, these arguments suggest that if the excess measured close
to the radio loud QSO sightlines is real, it is probably largely
associated with the QSOs, i.e. at $z_{\rm em}$. This result is
consistent with the observation that radio loud QSOs at $z\simeq0.5$
are frequently situated in cluster environments (cf. Yee \& Green 1987;
Ellingson, Yee \& Green 1991).

The physical scale over which the excess is seen might be understood
if  we are seeing luminous galaxies in the cores of distant groups or
proto-clusters centred around the QSO.  Since we would only be able to
detect the most luminous, concentrated components of such distant
systems, the excess seen would be both of limited significance at our
$K$ limit and  be compactly distributed.  To illustrate this effect at
the mean redshift of the radio loud objects ($\langle z_{\rm
em}\rangle=2.37$), we have taken  deep $K$-band images of rich clusters
of galaxies at $z=0.31$ from the study of Barger et al. (1995) and used
these to simulate the appearance of a non-evolving cluster observed
under the same conditions, and reaching the same depths as our QSO
images. We find that we would detect only 2--3 of the brightest cluster
galaxies inside a circle of $7\,$arcsec radius, with an average
luminosity of  $\langle M_K\rangle  \simeq -25.7$, i.e., about 0.5
magnitudes brighter than $L_K^{\ast}$. Figure~3 shows the absolute
magnitude distribution of all the galaxies found with a distance
$1<\theta<7\,$arcsec from the radio loud QSOs, assuming they are at
$z_{\rm em}$.  In total there are 16 objects, and we would expect
$\simeq 5$ from the field $K$ counts indicating an excess of $\simeq
11$ objects or 2.8 objects per radio loud QSO, with an average absolute
magnitude $\langle M_K\rangle  \simeq -25.9$.  The absolute magnitude
distribution of the excess objects observed ($M_K\lesssim -26.5$) is
$\sim 1\,$mag brighter than would be expected on the basis of
present-day brightest cluster galaxies, suggesting some luminosity
evolution from $z\sim2.4$ to $z\sim0$. Such evolution would not be
unreasonable over a $\simeq 10\,$Gyr period, depending on the star
formation history of the galaxies. Note, however, that without accurate correction for field contamination
there is some uncertainty in the intrinsic magnitude distribution of
the excess objects. Given the size of the sample, it is beyond the
scope of this paper to attempt building an ``uncontaminated''
luminosity function using the field galaxy number counts.

\begin{figure}
\psfig{figure=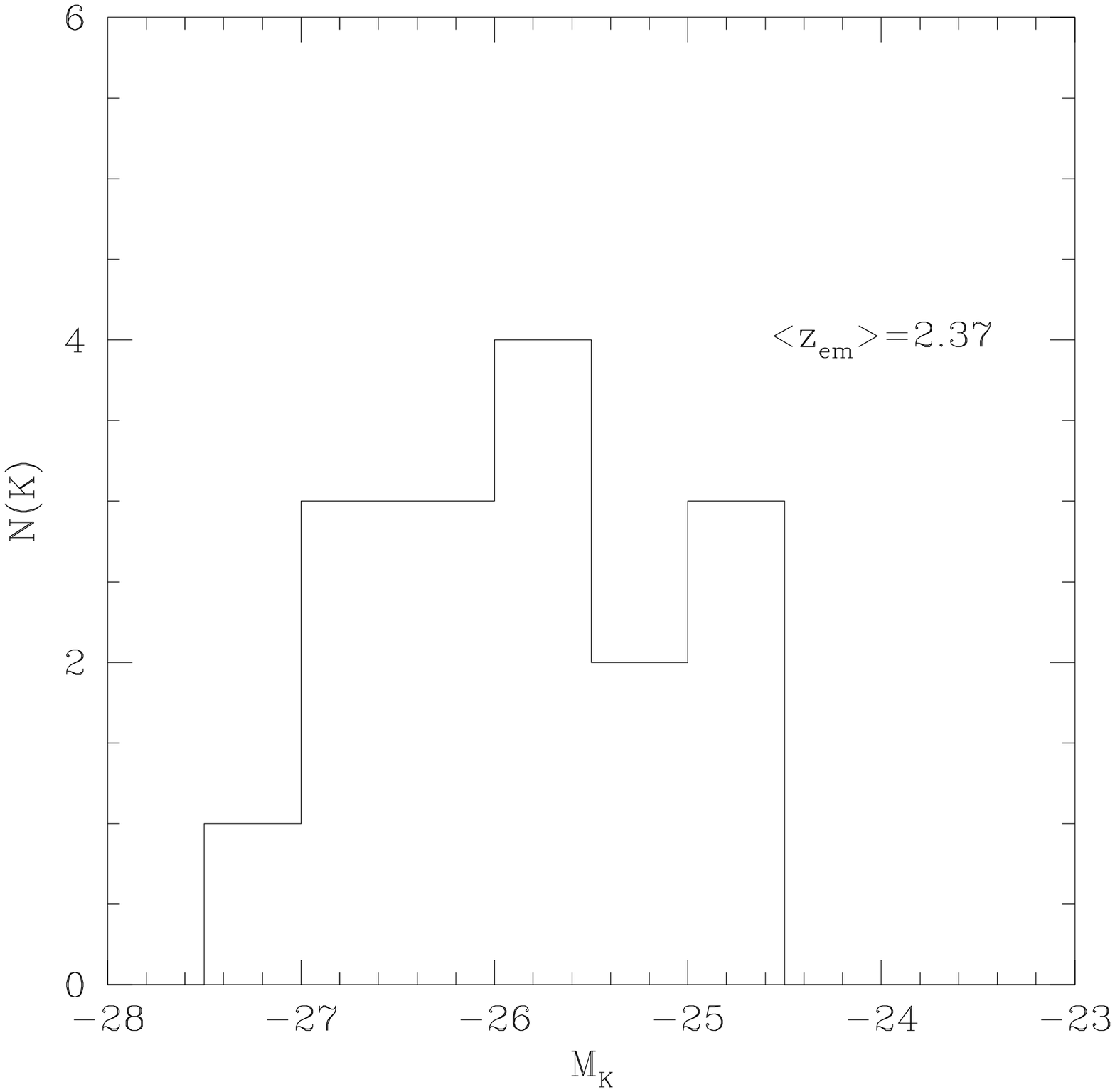,width=80mm,height=70mm}
\caption{Absolute magnitude distribution of all the galaxies found 
with a distance $1<r<7\,$arcsec from the radio loud QSOs, assuming 
they are at $z_{\rm em}$.}
\end{figure}

\subsection{Objects under the QSO PSF: New DLA galaxy candidates}

The primary aim of our observations was to search for candidate
galaxies close to the QSO sightline that might be responsible for DLA
absorption.  Our PSF subtraction procedure reveals galaxies very close
to the sightlines of Q0841+129 and Q1215+333, two of the radio-loud
objects.  Notwithstanding the excess population discussed earlier,
these are promising candidate absorbers (Figure~4).

\begin{figure}
\psfig{figure=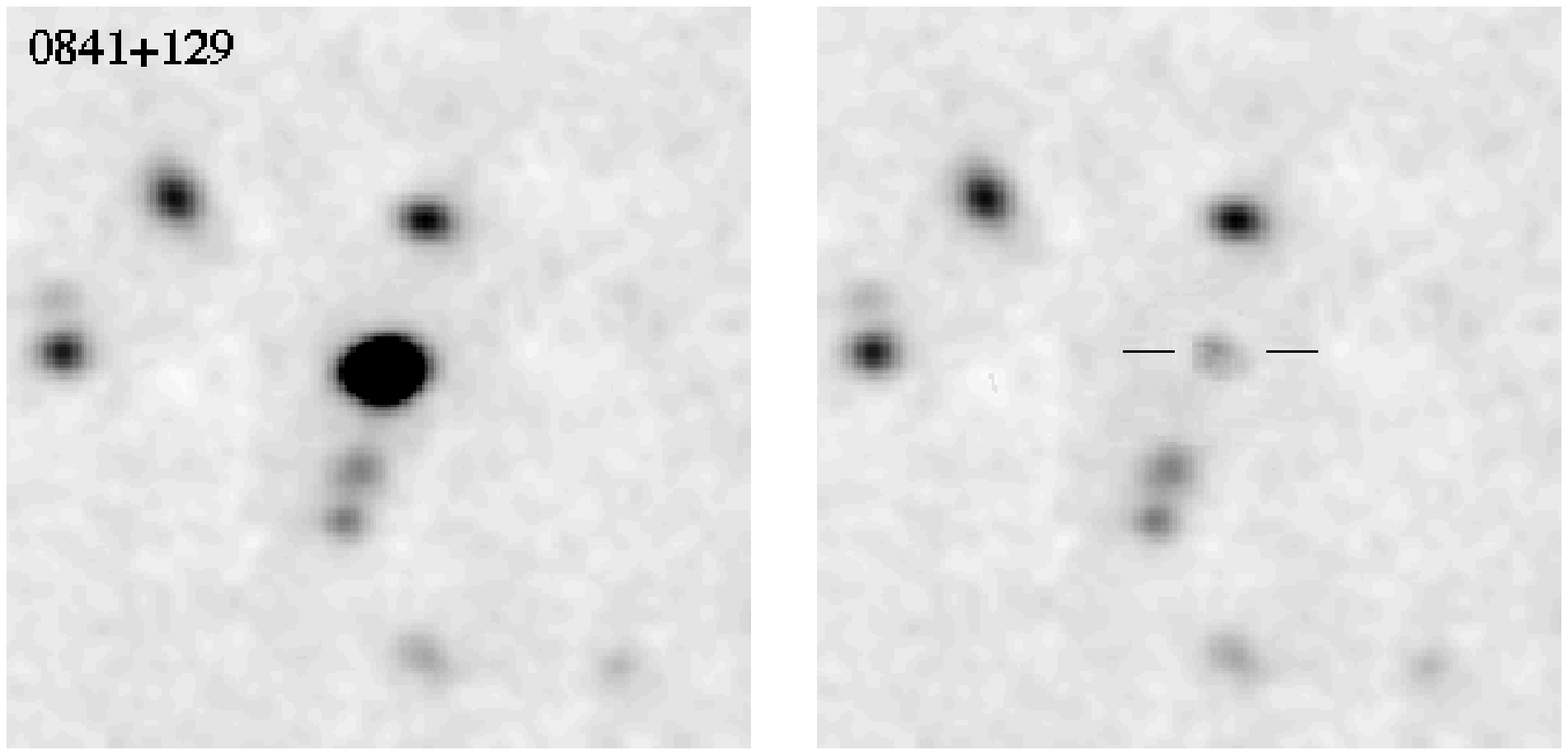,width=85mm}
\vskip 3mm
\psfig{figure=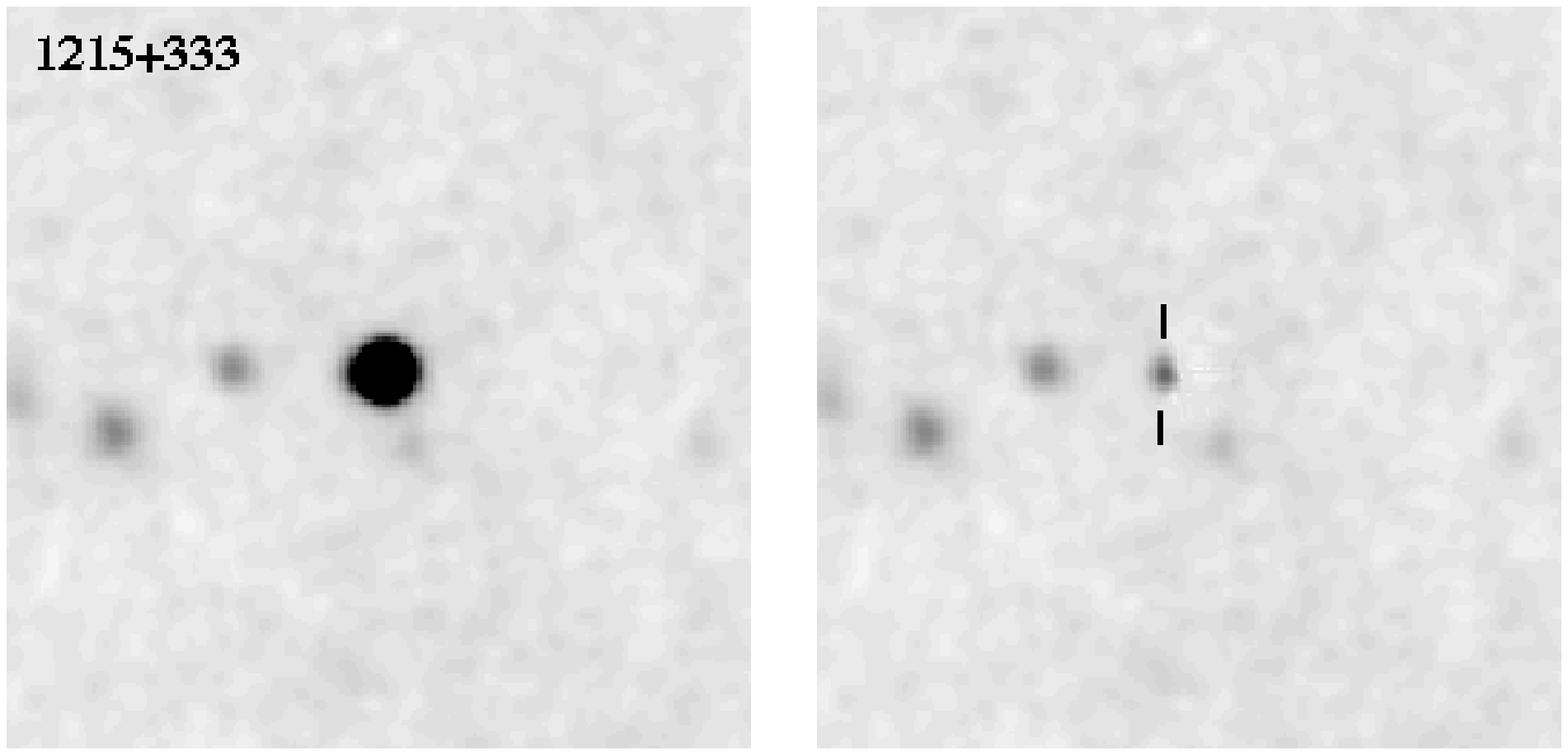,width=85mm}
\caption{Enlargement of the $K$-band images presented in Figure~1 for
0841+129 and 1215+333 showing the DLA candidate absorber galaxies.
Each frame is $30^{\prime\prime}\times30^{\prime\prime}$. }
\end{figure}

A $K=19.9$ galaxy was found $1.2^{\prime\prime}$ NW of the BL-Lac
object 0841+129 ($z_{\rm em}>2.51$), which has three DLA systems at
$z_{\rm abs}=1.8610$, $2.375$ and $2.477$ (Hazard, private
communication). At the average redshift of the DLAs $\langle z_{\rm
abs}\rangle=2.237$, the observed angular distance corresponds to an
impact parameter $\simeq 10\,$kpc. At that redshift, the galaxy would
be about $0.5\,$mag brighter than a present-day $L_K^{\ast}$ galaxy.

A $K=20.1$ galaxy appears $1.3^{\prime\prime}$ E of Q1215+333 ($z_{\rm
em}=2.606$), which has a DLA system at $z_{\rm abs}=1.999$\footnote{
This absorber was the subject of an infrared spectroscopic study by
Elston et al. (1991), who claimed a detection of 
[OII]$\lambda 3727$ and H$\beta$
emission at the absorber's redshift. This detection was not confirmed
by subsequent studies, and it is now believed to have been spurious
(Lowenthal 1995, private communication).}.  If the galaxy is at the
absorption redshift, its impact parameter would be $\simeq 11\,$kpc,
and its luminosity would be close to present-day $L_K^{\ast}$.

There are no other known metallic absorption line systems at redshifts
lower than that of the DLA systems in either of these two lines of
sight. Specifically, intermediate resolution spectroscopy is available
in the range $\lambda\lambda3330$--$4810$ ($2.5$\AA\ resolution) and
$\lambda\lambda3100$--$11000$ ($4$--$7$\AA\  resolution)
 for Q0841+129 (Hazard, private communication). For Q1215+333, Wolfe et
al. (1993) published good quality $1$\AA\ resolution spectroscopy in
the range $\lambda\lambda3500$--$4850$ and $2.5$\AA\ resolution
spectroscopy in the $\lambda\lambda6625$--$7325$ range which reveal no
lower redshift metallic absorption systems.

Given the average $K$-band field number counts, the limiting magnitudes
of the images, and $\theta_{\rm min}$ for the PSF subtraction, the
probability of finding two galaxies as close as the ones found here is
$\lesssim2$\% for the complete QSO sample making the two galaxies
promising candidates for the DLA absorption. Assuming this is the case,
their luminosities are comparable to those of the lower redshift
galaxies responsible for the metal line absorbers, but with smaller
impact parameters. This is certainly consistent with the hypothesis
that different classes of absorbers trace the same population of
galaxies with different cross sections. The DLA systems would then be
associated with the inner, denser regions, while metallic lines and
Lyman limit systems sample the more tenuous and extended haloes
(Steidel et al. 1994).

Of course, this conclusion is not yet convincing given the majority of
the DLA systems in our sample have no obvious visible counterpart.  At
lower redshift, Steidel et al. (1994) have imaged two DLA absorbers at
$z_{\rm abs}=0.3950$ and  $0.6922$ and found one to be a low surface
brightness galaxiy (which would certainly not be detected at our
redshifts)  whereas the other has $L>0.25\,L^{\ast}$.  This disparity
in galaxy properties, together with the difficulty of identifying
sources closer to the distant QSO sight lines than $\theta_{\rm min}$ may
explain the low success rate in finding luminous counterparts.

When the excess of sources found in the radio-loud QSO sightlines is
taken into account,  the probability of finding very close companions
to the QSOs ($\simeq8$\% for the whole survey) increases slightly. In
other words, the two DLA candidates  might be a subset drawn from the
clustered component around the radio-loud QSOs.  Obtaining redshifts is
the only means of solving these ambiguities.  Given the faintness of
the objects and their suspected redshift range, the best hope resides
in near-infrared spectroscopy, particularly if they are star-forming
galaxies whereupon the most likely emission lines would be redshifted
into the $JHK$ windows. The very high sky background in the near
infrared has been a major obstacle in such work thus far, but the
forthcoming availability  of purpose-built infrared spectrographs,
including those offering OH-suppression, may render such observations
feasible in the near future.

\section{Summary} 

We have obtained very deep ($K_{\rm lim}\simeq 21.3$--$21.7$) images in
good seeing conditions of the fields of 10 QSOs with damped
Lyman-$\alpha$ absorption systems in the $1.7<z_{\rm abs} <2.5$ range.
The main aim is to constrain the properties of  those galaxies
responsible for the DLA absorption and, more generally, to find samples
of high redshift galaxies for evolutionary studies. From the analysis
of those images, several results emerge:

1. For two of the QSOs, point spread function subtraction techniques
reveal the existence of one galaxy in each case very close to the QSO
line of sight (projected distance $1.2$--$1.3\,$arcsec, or
$\simeq10\,$kpc at the absorber redshift). We argue that, given the
small probability of a chance alignment and the implied impact
parameter at $z_{\rm abs}$,  these two objects are promising candidates for
the DLA absorption.  Both would have luminosities close to $L_K^{\ast}$
and their properties would be consistent with trends identified in
metal line absorber galaxies found at lower redshifts.

2. The radial distribution of faint sources around the QSOs shows good
agreement with that expected from published $K$-band number counts for
projected distances $\gtrsim7\,$arcsec.  At $\theta\lesssim7\,$arcsec,
radio quiet QSOs show no sign of excess in the number of objects, but
the radio loud QSOs show an excess of $2.8$ sources per line of sight
significant at the $99.5$\% level.

3. Recognising that the statistical significance of the excess is not
very high given the number of lines of sight sampled,  the difference
in behaviour detected for the radio quiet and radio loud QSOs is
nonetheless suggestive.  Since the DLA systems should not be affected
by the radio properties of the QSOs ($z_{\rm abs}\ll z_{\rm em}$),  we
suggest  the excess may be associated with the QSOs themselves. If that
is the case, radio loud QSOs at $z\simeq2.4$ could well reside in
groups or clusters of galaxies much in the same way as closer
counterparts, and we are just seeing the 2--3 brightest galaxies
there.

4. The average luminosity of the excess galaxies, if placed at $z_{\rm
em}$, is $\simeq 2L_K^{\ast}$, and the magnitude distribution reaches
$\simeq 2\,$mag brighter than $L_K^{\ast}$. If these galaxies are the
high redshift counterparts of present day brightest cluster galaxies,
it implies luminosity evolution in the $K$-band of $\sim1\,$mag from
$z\simeq2.4$ to the present, in the sense that high-$z$ galaxies were
brighter than today.

Imaging high redshift QSO sightlines is an efficient method to identify
likely high-$z$ galaxies for evolutionary studies.  To the precision
where evolutionary comparison with lower redshift galaxies are
concerned, the small difference in look-back-time makes it immaterial
whether the galaxies are at $z_{\rm abs}$ or at $z_{\rm em}$.  A larger
sample of radio loud and radio quiet QSOs, with and without DLA systems
is needed in order to add weight to our statistical conclusions.  Once
the excess is firmly established,  HST imaging would provide morphology
and colours and be useful in searching for objects even closer to the
QSO line of sight.  Obtaining redshifts can resolve the ambiguities and
this may be practical shortly via the use of  near-infrared
spectrographs offering OH-suppression where faint emission lines may be
seen against the contrast of the strong infrared background .

\section*{Acknowledgements}

We would like to thank UKIRT staff for their assistance in securing
these observations, and  Guinevere Kauffman, James Lowenthal and Simon
White for very useful discussions.  We are also grateful to Mark
Dickinson, the referee of this paper, for his comments.  AAS
acknowledges generous financial support from the Royal Society.

\end{document}